# Criticisms on and Comparison of Experimental Channel Backscattering Extraction Methods

Gino Giusi, Felice Crupi, Paolo Magnone

DEIS, University of Calabria, Via P. Bucci 41C, I-87036 Arcavacata di Rende (CS), Italy

## Abstract

In this paper we critically review and compare experimental methods, based on the Lundstrom model, to extract the channel backscattering ratio in nano MOSFETs. Basically two experimental methods are currently used, the most common of them is based on the measurement of the saturation drain current at different temperatures. We show that this method is affected by very poor assumptions and that the extracted backscattering ratio strongly underestimates its actual value posing particular attention to the backscattering actually extracted in nano devices. The second method is based on the direct measurement of the inversion charge by CV characteristics and gets closer to the physics of the backscattering model. We show, through measurements in high mobility p-germanium devices, how the temperature based method gives the same result of the CV based method once that its approximations are removed. Moreover we show that the CV based method uses a number of approximations which are partially inconsistent with the model. In particular we show, with the aid of 2D quantum corrected device simulations, that the value of the barrier lowering obtained through the CV based method is totally inconsistent with the barrier lowering used to correct the inversion charge and that the extracted saturation inversion charge is underestimated.

# I. Introduction

Due to the continuous downscaling of MOSFET geometry, improved physical models are needed to accurately study the charge transport in the channel. One of the simplest and most successful models was proposed by M. Lundstrom [1-4] built on the Natori model for ballistic MOSFETs [5]. In his *Backscattering* model, the charge transport in the channel is regulated by the injection of the near equilibrium thermal carriers at the top of the source-channel potential barrier (the virtual source). Only a fraction of the injected carriers reaches the drain side due to scattering in the channel. In saturation the backscattering coefficient $r$ is defined as the ratio between the backward-going flux ($\Gamma^-$) and the forward-going flux ($\Gamma^+$) at the virtual source (Fig. 1).

The strength of the model is that it provides just a number, the backscattering coefficient $r$, which includes all scattering mechanisms in the channel. Quasi-ballistic transistors have $r$ close to zero [6], so that all the injected carriers reach the drain side providing maximum current drive. Technology developers and transistor designers must aim at devices with low $r$ in order to enhance performance. In this sense the backscattering coefficient is a parameter which provides information about the advantage, in terms of ON current, related to the scaling of a given technology (material and/or architecture). For these reasons it is very important to have experimental procedures which accurately estimate the backscattering coefficient [23]. Up to now, two methods have been used to extract $r$: a temperature-based method proposed by Chen [7], in which the backscattering coefficient is extracted by measuring the saturation current at different temperatures, and a capacitance-voltage (CV) based method proposed by Lochtefeld [8-10], in which the inversion charge is obtained from the measurement of the gate-to-channel capacitance. Both methods are based on approximations that can lead to sensible errors in the backscattering estimation. Especially the first, which is more used due to its simplicity [11-15], is affected by a number of approximations resulting in an excessive underestimation of $r$. In this paper we compare the two methods and address their approximations. We show how the temperature based method gives the same result of the CV based method once that its approximations are removed. Moreover we show that the CV based method uses itself a number of approximations which are partially inconsistent with the model. In particular we show, with the aid of 2D quantum corrected device simulations, that the value of the barrier lowering obtained through the CV based method is totally inconsistent with the barrier lowering used to calculate the saturation inversion charge and that the extracted saturation inversion charge is underestimated.

The remainder of the paper is divided as stated in the following. In Section II the backscattering model is presented. In Section III the temperature based method and the CV based method are presented. In Section IV and in the Appendix the methods are compared through measurements in short channel germanium pMOSFETs. In Section V the issues related to the Lochtefeld method are discussed with the aid of two-dimensional device simulation. In Section VI we present our conclusions.

# II. The Channel Backscattering Model

Due to its continuity along the channel, the saturation drain current can be evaluated at the top of the source-channel potential barrier (the virtual source $x_0$ in Fig. 1) where carriers fill $k$-states according to semi-maxwellian distributions [3, 24]

$$I_{D,sat} = WQ_{sat}v = WQ_{sat}v_{inj}b \;, \qquad b = \frac{1-r}{1+r} \qquad r = \frac{I^-}{I^+} \tag{1}$$

where $I_{D,sat}$ is the saturation drain current, $W$ is the device width, $Q_{sat}$ is the inversion charge per unit of area in saturation, $v$ is the average carrier velocity, $v_{inj}$ is the injection velocity, $b$ is the ballisticity coefficient, $r$ is the backscattering coefficient, $I^-$ is the negative directed current and $I^+$ is the positive directed current. All quantities in Eq. 1 are evaluated at the virtual source. $v$ is also called the *effective injection velocity* because it is the actual velocity of the injected carriers. Carriers at the semiconductor-oxide interface are confined in a triangular well so that they occupy discrete energy levels (sub-bands). For each sub-band we can calculate an inversion charge, an injection velocity and a resulting current given as in Eq.1. Total current is the sum of the contributions of each sub-band. However, due to the high transverse field, often only one sub-band is populated. Because carriers at the top of the barrier are in thermal equilibrium, $v_{inj}$ is close to the average 1D thermal velocity $v_{th}$. In operating conditions (above threshold), MOSFETs work in degenerate conditions and the injection velocity is different from $v_{th}$. If the one sub-band approximation holds, the injection velocity is related to the thermal velocity by

$$v_{inj} = v_{th}\frac{F_{1/2}(\eta)}{F_0(\eta)} \qquad v_{th} = \sqrt{\frac{2kTm_c}{\pi m_{DOS}^2}} \qquad \eta = \pm\frac{E_{FS}-E_1}{kT} \tag{2}$$

where $k$ is the Boltzmann constant, $T$ is the absolute temperature, $m_c$ is the conduction effective mass of the considered sub-band, $m_{DOS}$ is the density of states effective mass of the considered sub-band, $F_{1/2}$ is the Fermi-Dirac integral of order ½, $F_0$ is the zero-order Fermi-Dirac integral and $\eta$ is the energy distance, in unit of $kT$, of the populated sub-band (with energy $E_l$) with respect to the source quasi Fermi level $E_{FS}$ (the plus sign in Eq. 2 is for electrons, the minus sign is for holes). The inversion charge at the virtual source with a generic drain to source voltage ($V_D$) applied is related to the backscattering coefficient by [4]

$$Q_{inv} = qN_{2D}\frac{[1+r]}{2}F_0(\eta)\left[1+\left(\frac{1-r}{1+r}\right)\frac{F_0\left(\eta-\frac{qV_D}{kT}\right)}{F_0(\eta)}\right] \qquad N_{2D} = kTg\frac{m_{DOS}}{\pi\hbar^2} \qquad (3.1)$$

where $q$ is the electron charge, $N_{2D}$ is the two dimensional density of states, $\hbar$ is the reduced Plank constant, $g$ is the sub-band degeneracy. Let us note that in Eq. 3.1 $Q_{inv}$ is calculated through the 2D density of states because the charge is confined along the transverse direction so that carriers are free to move only in two dimensions. In equilibrium ($V_D$=0) Eq. 3.1 becomes

$$Q_{eq} = Q_{inv}(V_D = 0) = qN_{2D}F_0(\eta_0) \qquad (3.2)$$

while in saturation, where $V_D >> kT/q$

$$Q_{sat} = Q_{inv}\left(V_{D,sat}\right) \cong qN_{2D}\frac{[1+r]}{2}F_0(\eta) \qquad (3.3)$$

When the MOSFET is biased in the saturation region, Eq. 3.3 applies. But when the inversion charge is evaluated directly from a CV measurement, where $V_D$=0, Eq. 3.2 should be used. Equations 3 are very important because they provide a link between the inversion charge, the backscattering coefficient and the normalized potential ($\eta$).

# III. Extraction Procedures for Channel Backscattering

In this Section we discuss the two channel backscattering extraction methods commonly used with their approximations: the Chen method and the Lochtefeld method. The first method is much more used [11-15] probably because it needs just the measurement of the drain current at different temperatures, while the second method requires the measurement of the drain current and the measurement of the gate to channel capacitance (only at the desired temperature). Despite it is simpler, the Chen method makes use of a number of approximations which leads to a significant underestimation of the backscattering coefficient. The Lochtefeld method instead produces more accurate results but it is also affected by a number of inconsistencies with the backscattering model which are discussed in Section V.

## IIIA. The Chen Method

The temperature-based method for the extraction of the backscattering coefficient was proposed for the first time by Chen et al. [7]. It exploits the temperature dependence of the drain current. The assumptions of the method are

A1. $Q_{sat}(x_0) \approx C_{eff}\left(V_{GS} - V_{T,sat}\right)$

A2. no carrier degeneration, so that $v_{inj} \cong v_{th}$

A3. $r$ depends on the temperature $T$ as $r(T) = \frac{1}{1+\frac{c}{T^2}}$ , from which $\frac{\partial r}{\partial T} = \frac{2r(1-r)}{T}$ where $c$ is a constant, $C_{eff}$ is a constant effective oxide capacitance, $V_{T,sat}$ is the saturation threshold voltage. Indeed A3 comes from two other approximations, that is from the assumed dependence of the low-field mobility $\mu^0$ on the temperature (A4) and from the independence of the longitudinal electric field $\varepsilon$ at the top of the barrier from the temperature $\varepsilon(x_0^+) = \frac{kT}{q}/l$ (A5), where $l$ is the length of the critical layer ($kT$). In fact, neglecting degeneration, the mean free path ($\lambda$) can be related to the low field mobility ($\mu^0$) by [4, 16]

$$\frac{\lambda}{l} = \frac{2\mu^0 kT/q}{l\, v_{inj}} \qquad (4)$$

By using the assumptions

A4. $\mu^0 \propto T^{-1.5}$

A5. $\frac{\partial \varepsilon(x_0^+)}{\partial T} = 0$

we obtain

$$\frac{\lambda}{l} = \frac{2\mu^0 kT/q}{l\, v_{inj}} = \frac{2\mu^0 \varepsilon(x_0^+)}{v_{inj}} \propto \frac{T^{-1.5}}{\sqrt{T}} = T^{-2} \text{ so that } r(T) = \frac{1}{1+\frac{\lambda}{l}} = \frac{1}{1+\frac{c}{T^2}} \text{ which is A3.} \qquad (5)$$

By using A1 and A2, Eq.1 becomes

$$I_{D,sat} \approx W C_{eff}\left(V_{GS} - V_{T,sat}\right) v_{th} \frac{1-r}{1+r} \qquad (6)$$

By differentiating the drain current with respect to the temperature one obtains

$$\frac{\partial I_{D,sat}}{\partial T} = I_{D,sat}\left[\frac{1}{2T} - \frac{2}{1-r^2}\frac{\partial r}{\partial T} - \frac{\gamma}{V_{GS}-V_{T,sat}}\right] = I_{D,sat}\alpha \qquad (7)$$

where $\gamma$ is the temperature coefficient of $V_{T,sat}$, i.e. $\gamma = \partial V_{T,sat}/\partial T$ and $\alpha$ is the term inside the brackets. So by fitting $\alpha$ and $\gamma$ and extracting $I_{D,sat}$ from IV measurements at different temperatures and by using A3 one can extract the backscattering coefficient from Eq. 7.

It appears clear that A4 is quite arbitrary because the dependence of the mobility with the temperature is function of the material, the doping level, the strain, and so on. Also A5 appears quite arbitrary. It could be justified considering that the length of the critical $l$ layer is so short to assume a linear shape of the potential inside this region. Differently from the argumentation of Zilli *et al.* [17] which attributes to A1 the main approximation, we will found in Section IV that A5 has the main impact on the extracted backscattering.

## IIIB. The Lochtefeld Method

The Lochtefeld method is based on the extraction of the backscattering coefficient by directly measuring the physical quantities that appear in Eq. 1. Drain current is measured with a current-voltage (IV) measurement, while $Q_{inv}$ by integrating the gate to channel capacitance up to the applied gate voltage. Because of the difficulty in measuring the inversion charge in short channel devices, $Q_{inv}$ is first measured in a longer reference device, and after it is corrected for the $V_T$ roll-off and for the DIBL by using the maximum transconductance method and the constant current threshold method in the subthreshold region, respectively. A further correction is needed to take into account for the series resistance. This was the way used by Lochtefeld *et al.* [8, 9] to measure the effective injection velocity in short channel devices. Following this approach Dobbie *et al.* [10] extracted the channel backscattering ratio in germanium pMOSFETs by using Eq. 1, Eq. 2 and Eq. 3.2.

Note that the use of the equilibrium inversion charge (Eq. 3.2) is not appropriate because in the presence of transport ($V_D \neq 0$) the saturation inversion charge should be used (Eq.3.3). In the same way only Eq. 3.3, and not Eq. 3.2, provides the right link between inversion charge and the normalized potential $\eta$ when $V_D \neq 0$.

## IV. Methods Comparison through Measurements in Germanium p-MOSFETs

In this subsection we compare the two previously discussed methods for the extraction of the backscattering coefficient in advanced high mobility p-germanium devices [18-20]. In particular we remove one by one the approximations of the Chen method through the procedure discussed in Appendix, and we see how the final result is very close to that of the Lochtefeld method. In Fig. 2 we plot the backscattering coefficient (empty symbols) in germanium pMOSFETs for different gate lengths extracted with the Lochtefeld method and with the Chen method at the bias $V_G$=$V_D$=-1V. The gate stack consists of a silicon passivation layer of 0.4 nm, 4 nm of $HfO_2$, and of a TiN/TaN metal gate. Details about the devices can be found in [20]. Schrodinger-Poisson simulations show that the one sub-band approximation holds for the applied bias so that the backscattering model can be used.. Fig. 2 also shows mobility data for different channel lengths calculated by a simple integration of the gate to channel capacitance taking into account for the $V_T$ roll-off and the series resistance which has been extracted by a common linear extrapolation technique [22]. The degraded mobility for shorter gate lengths indicates a higher scattering rate. Anyway as the gate length is reduced the backscattering is expected to be lower because of the reduced region where scattering is important (the critical layer) [1]. The value of the backscattering obtained with the Chen method appears to underestimate strongly the value obtained with the Lochtefeld method. Moreover in Fig. 2 the values of the backscattering coefficient obtained with the Chen method, by removing one by one its approximations, are also shown. Carrier degeneration (A2) and inversion charge estimation (A1) would decrease the value of the backscattering while the dependence of the mobility (A4) and of the electric field (A5) upon the temperature raise its value. It can be noticed that the stronger approximation appears to be the assumed independence of the longitudinal electric field with the temperature (A5). In fact when A1, A2, A4 are removed the results of the two methods continue to be strongly different, while when A5 is also removed the results of the two methods appear very close one to the other. In fact in this case the two methods are totally equivalent as discussed in the Appendix. The little difference is due to the error in extracting the derivatives with respect to the temperature. We want to point out that the most recent papers [11-15] use the Chen method to extract the backscattering coefficient and that the obtained results can be biased according to our previous considerations. Moreover, Zilli [17] attributed to A1 the major impact on the extracted backscattering, while here we found that is A5 the poorer approximation.

# V. Issues in the Lochtefeld Method

In this Section we discuss the approximations and inconsistencies connected to the Lochtefeld method with the aid of 2D device simulations performed through Medici [21]. Other than the Fermi-Dirac statistics and the low field (ARORA) and surface scattering (PRPMOB) mobility models, density gradient quantum correction has been turned on over the classical drift diffusion scheme. Simulated devices are silicon n-MOSFETs with $SiO_2$ gate oxide, polysilicon gate, bulk doping of $10^{18}$ $cm^{-3}$, oxide thickness of 1.2 nm and different gate lengths down to 70 nm. A fundamental point is related to the correct extraction of the saturation inversion charge. As stated in Sections IIIB, it is difficult to measure the saturation charge in short channel devices because of parasitic capacitances (overlap and instrumentation). The usual strategy is that of measuring the inversion charge by a CV in a longer channel device in the way that the measurement of the higher device capacitance is not affected by parasites. This equilibrium charge is corrected for the $V_T$ roll-off in order to obtain the equilibrium charge in the considered short channel device. The saturation charge is now obtained correcting the equilibrium charge with the DIBL which is measured in the sub-threshold regime. There are two mistakes in this procedure. The first one is connected to the charge obtained by the CV measurement. To understand this point let us observe Fig. 3 where it is shown the inversion charge at the peak point along the transverse direction (1nm far from the interface) in strong inversion in the long reference device (L=10 μm) and in the short channel device (L=100 nm) as it appears during a CV measurement, that is with the drain voltage ($V_D$) equal to zero. The charge extracted by a CV measurement is strongly influenced by the charge variation close to the drain and source regions in the short channel device, making its measurement very difficult, while in the long channel device these regions have poor influence because the extracted charge is averaged along the long channel so that the measured charge in the long channel corresponds to the plateau as indicated in the Fig. 3. The equilibrium charge which corresponds to the short channel device is obtained taking into account for the $V_T$ roll-off. However the needed charge to be used in Eq. 1 is the charge at the virtual source in saturation. In Fig. 3 it is also shown the saturation inversion charge density in the short channel device and the value needed in Eq. 1 is indicated. As can be noticed, the position of the virtual source is very close to the source edge where the charge density profile is strongly affected by the lateral profile of the source doping. In the Lochtefeld method, this value is calculated by correcting the inversion charge with a DIBL extracted in the sub-threshold regime. As can noticed in Fig. 4a), the DIBL is indeed a strong function of the bias point and usually it is higher in the sub-threshold regime with respect to the inversion regime leading to an error in the inversion charge extraction. Anyway, because the lateral source doping profile can have an arbitrary shape, also if one has a method to calculate the DIBL at the operative bias point in inversion, we cannot assume that the charge calculated with this correction should reproduce perfectly the charge at the virtual source. The second point is related to the consistency of the equations used by the Lochtefeld method with respect to the model equations (1-3). As already stated in Section IIIB, the Lochtefeld method uses Eq. 3.2 to link the saturation inversion charge, calculated with the above discussed DIBL correction, and the normalized potential $\eta$. Indeed this equation describes the relationship between charge and potential in the case of equilibrium ($V_D$=0) where $r$=1 while the correct relationship to be used in saturation is Eq. 3.3. The two above discussed inconsistencies of the Lochtefeld method appear evident when one wants to compare the DIBL used to calculate the inversion charge (DIBL in sub-threshold) and the barrier lowering calculated as $kT/q{\cdot}\Delta\eta$, where $\Delta\eta=\eta-\eta_0$ and $\eta_0$ is the equilibrium normalized potential calculated from Eq. 3.2. From Fig. 4a) it is clear that these two quantities are strongly different and that, anyway, the barrier lowering calculated as $kT/q{\cdot}\Delta\eta$ is significantly lower with respect to the actual DIBL at the operative bias point in inversion. The same kind of comparison is plotted in Fig. 4b) for different gate lengths. In Fig. 5, the simulated expected saturation charge and the simulated charge extracted with the Lochtefeld method are compared. The charge obtained from the Lochtefeld method is affected by the sum of two inconsistencies as mentioned above. First, the charge is calculated with a DIBL higher with respect to the actual DIBL, that is with a DIBL in the sub-threshold regime instead of a DIBL in the inversion regime. This error should produce a charge higher with respect to the expected charge. Second, Eq. 3.2 is used in the Lochtefed method instead of Eq. 3.3, that is the increase in the charge from equilibrium to saturation is considered as due to only an electrostatic effect and scattering is neglected (the term $(1+r)/2$ in Eq. 3.3). Because scattering lowers the charge ( $(1+r)/2<1$ ) the value obtained with the Lochtefeld method appears lower with respect to the expected value. As a direct consequence of the above discussed issues, the value of the backscattering coefficient is affected.

# VI. Conclusion

In this paper we have revised and discussed the issues related to the principal methods used in the literature to extract the channel backscattering ratio in nano MOSFETs based on the Lundstrom transport model. We have shown that the Chen method, which is actually the most used, is affected by a number of strong approximations which lead to an underestimation of the extracted backscattering. In particular we found that the stronger approximation is that of assuming a linear shape of the potential inside the critical layer. The method has been compared to the Lochtefeld method, which better captures the physics of the backscattering model, through measurements in high mobility germanium p-MOSFETs. Moreover we discussed, with the aid of 2D quantum corrected device simulations, the inconsistencies of the Lochtefeld method which are mainly related to the saturation inversion charge calculation. We found that the barrier lowering used to calculate the charge is strongly different from the barrier lowering obtained as the result from the method and that the charge calculation is done neglecting the contribution of scattering. These issues, together with a difficult estimation of the charge at the virtual source, lead in erroneous calculation of backscattering and inversion charge.

# Appendix

In this Section we address the approximations used in the Chen method and we develop a new version almost free of approximations. A1 and A2 can be removed by measuring the inversion charge by a CV curve and calculating $\eta$ through Eq. 3.2 (note that here we use Eq. 3.2 instead of the correct Eq. 3.3 because we want to compare the temperature method with the standard CV method as they were proposed). Of course the measured charge must be corrected for short channel effects in the same way as in the Lochtefeld method. In this way the method looses the advantage of its simplicity with respect to the Lochtefeld method. A4 can be removed by calculating the actual mobility with the $g_D$ method [22]. To remove A5, the dependence of the longitudinal electric field with the temperature can be calculated by the Lochtefeld method. At this point the two methods become totally equivalent one to the other. Putting all together, we can calculate the derivative of the Eq.1 with respect to the temperature

$$\frac{1}{I_{D,sat}}\frac{\partial I_{D,sat}}{\partial T} = \frac{1}{Q_{sat}}\frac{\partial Q_{sat}}{\partial T} + \frac{1}{v_{inj}}\frac{\partial v_{inj}}{\partial T} + \frac{1}{b}\frac{\partial b}{\partial T} \tag{8}$$

The relative derivative of the current and the relative derivative of the inversion charge can be calculated by the measured current and inversion charge, the relative derivative of the injection velocity by Eq. 2, while the relative derivative of the ballisticity is related to the backscattering coefficient through

$$\frac{1}{b}\frac{\partial b}{\partial T} = \frac{2r}{1+r}\left[\frac{1}{\lambda}\frac{\partial \lambda}{\partial T} - \frac{1}{l}\frac{\partial l}{\partial T}\right] = \frac{2r}{1+r}\left[\frac{1}{\lambda}\frac{\partial \lambda}{\partial T} - \frac{1}{T} + \frac{1}{\varepsilon}\frac{\partial \varepsilon}{\partial T}\right] \tag{9}$$

which can be obtained by taking the relationships between $b$, $r$, $\lambda$, $l$ and $\varepsilon$ (Eq. 1, 4, 5). In Eq. 9, $\lambda$ is calculated by Eq. 4, while the electric field $\varepsilon$ from the Lochtefeld method. Equations 8 and 9 together can be considered a temperature version of the backscattering model.

# Figures Captions

Figure 1

Sub-band energy ($E_1$) profile in saturation along the channel direction $x$. The backscattering is defined as the ratio of the negative directed current ($I^-$) to the positive directed current ($I^+$) evaluated at the virtual source ($x_0$) which is the $x$ position corresponding to the maximum of the energy.

Figure 2

Experimental backscattering coefficient (empty symbols) and mobility (filled symbols) as function of the gate length in germanium pMOSFETs [20]. The backscattering has been calculated with the Lochtefeld method and with the Chen method. The values of the backscattering coefficient obtained with the Chen method by removing one by one its approximations are also shown. A1 and A2 will decrease the value of $r$ while A4 and A5 raise its value. It appears clear that A5 is the poorer approximation. When all approximations are removed it can be seen that the final result is very close to that of the Lochtefeld method.

Figure 3

Simulated inversion carrier concentration, at its peak value along the transversal direction (1nm far from the interface), in silicon nMOSFETs for the case of 1) the reference device (10µm) with $V_D$=0, 2) the device under test (100 nm) with $V_D$=0 and 3) the device under test in saturation ($V_D$=1V). It can be noticed that the value of the inversion charge per unit of area extracted by the CV on the longer channel device must be corrected, to the level of the inversion charge at the top of the barrier when $V_D$=1V in the short channel.

Figure 4

The barrier lowering simulated as a function of the gate bias (a) and as a function of channel length (b) with the Lochtefeld method and the one expected, calculated by taking the difference of the potentials at the virtual source, at the peak of charge along the transverse direction, for the cases $V_D$=1V and $V_D$=50mV . The simulated devices are silicon n-MOSFETs with poly-Si gate, bulk doping of $10^{18}$ $cm^{-3}$, oxide thickness $t_{ox}$=1.2nm. The expected DIBL is a strong function of the bias point. The DIBL calculated with the Lochtefeld procedure underestimates strongly the expected value showing a strong inconsistence with the sub-threshold DIBL used to calculate the inversion charge.

Figure 5

Inversion charge extracted with the Lochtefeld method and the expected value. The value obtained with the Lochtefeld method is affected by two inconsistencies: the wrong DIBL correction and the assumption of $r_{sat}$=1 in the charge calculation (Eq. 3.2).

# Figures

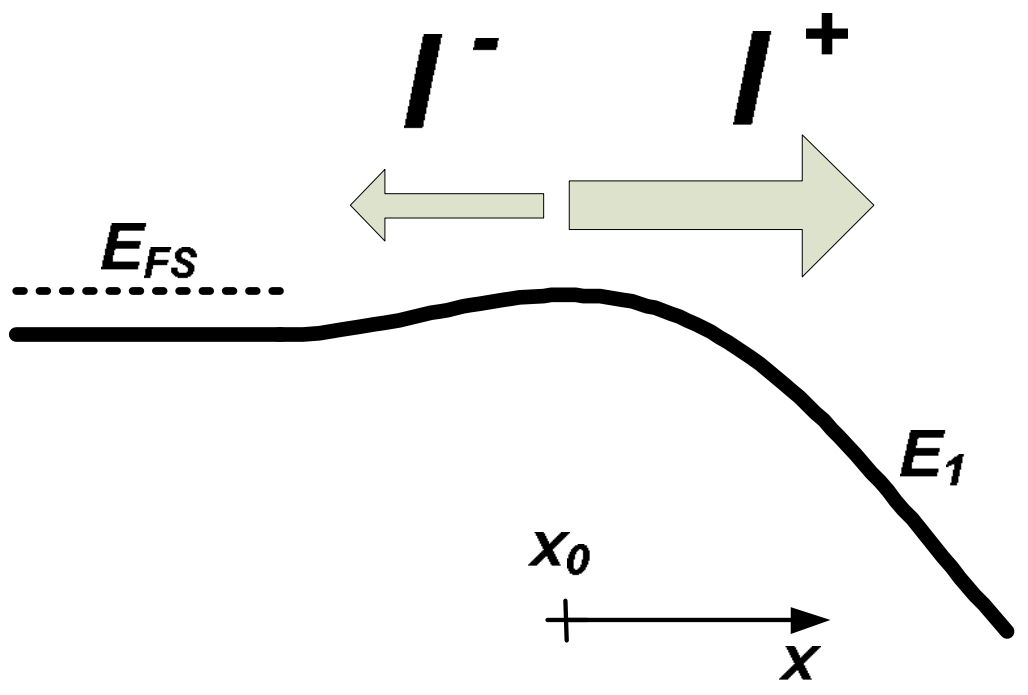

Figure 1

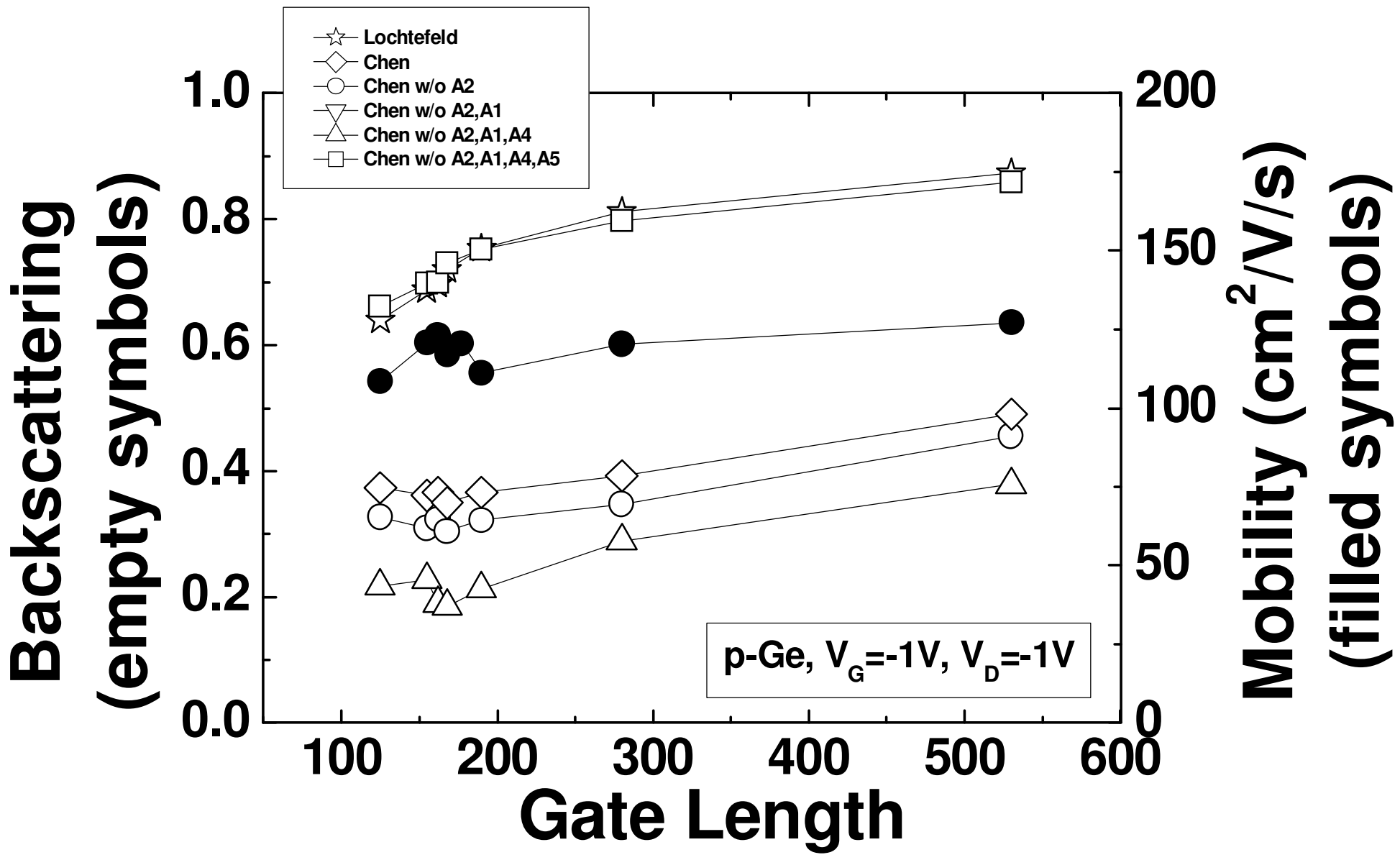

Figure 2

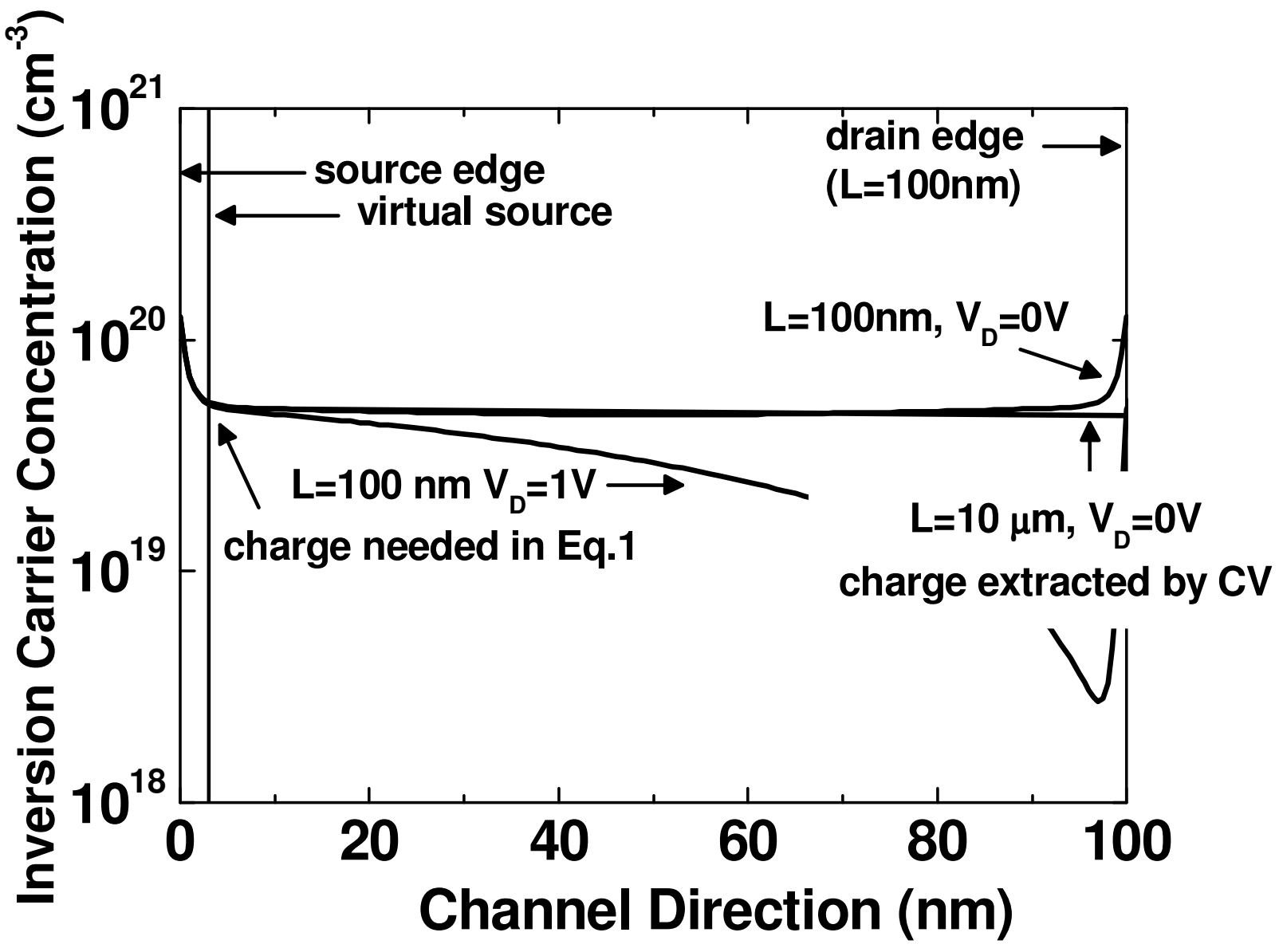

Figure 3

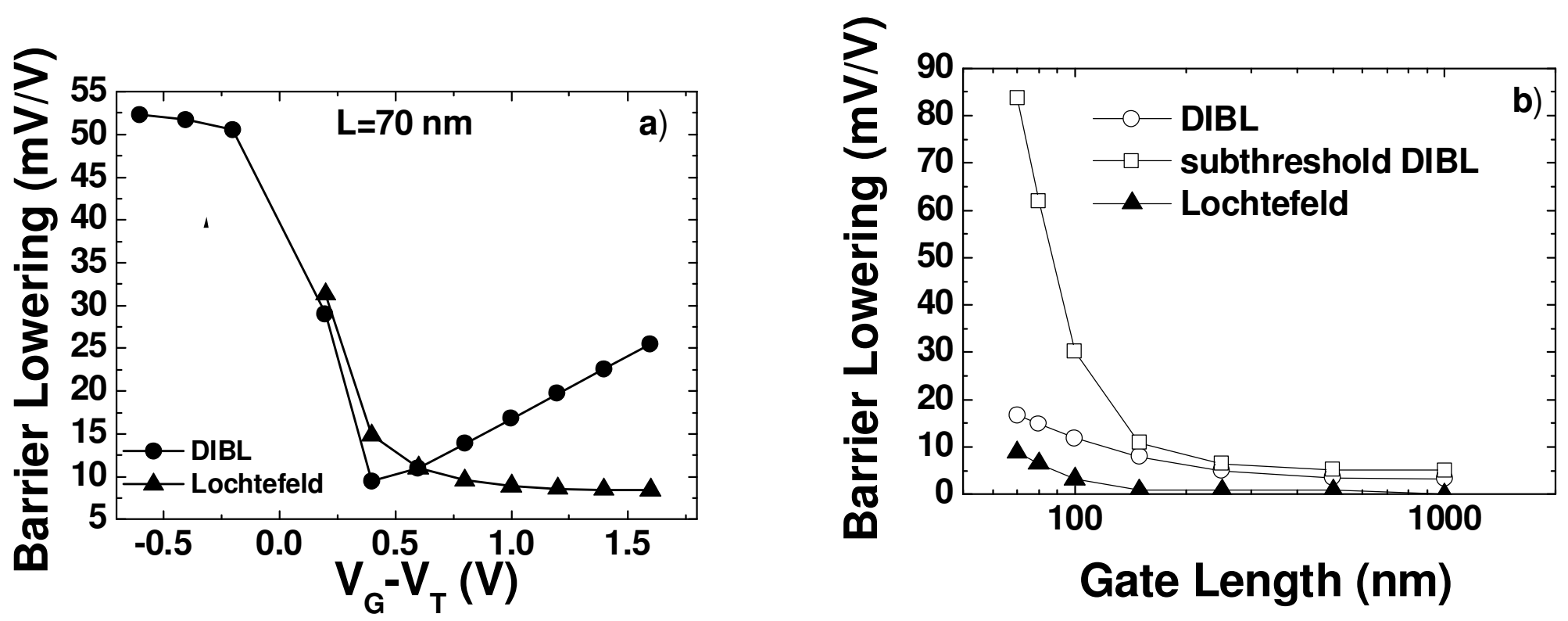

Figure 4

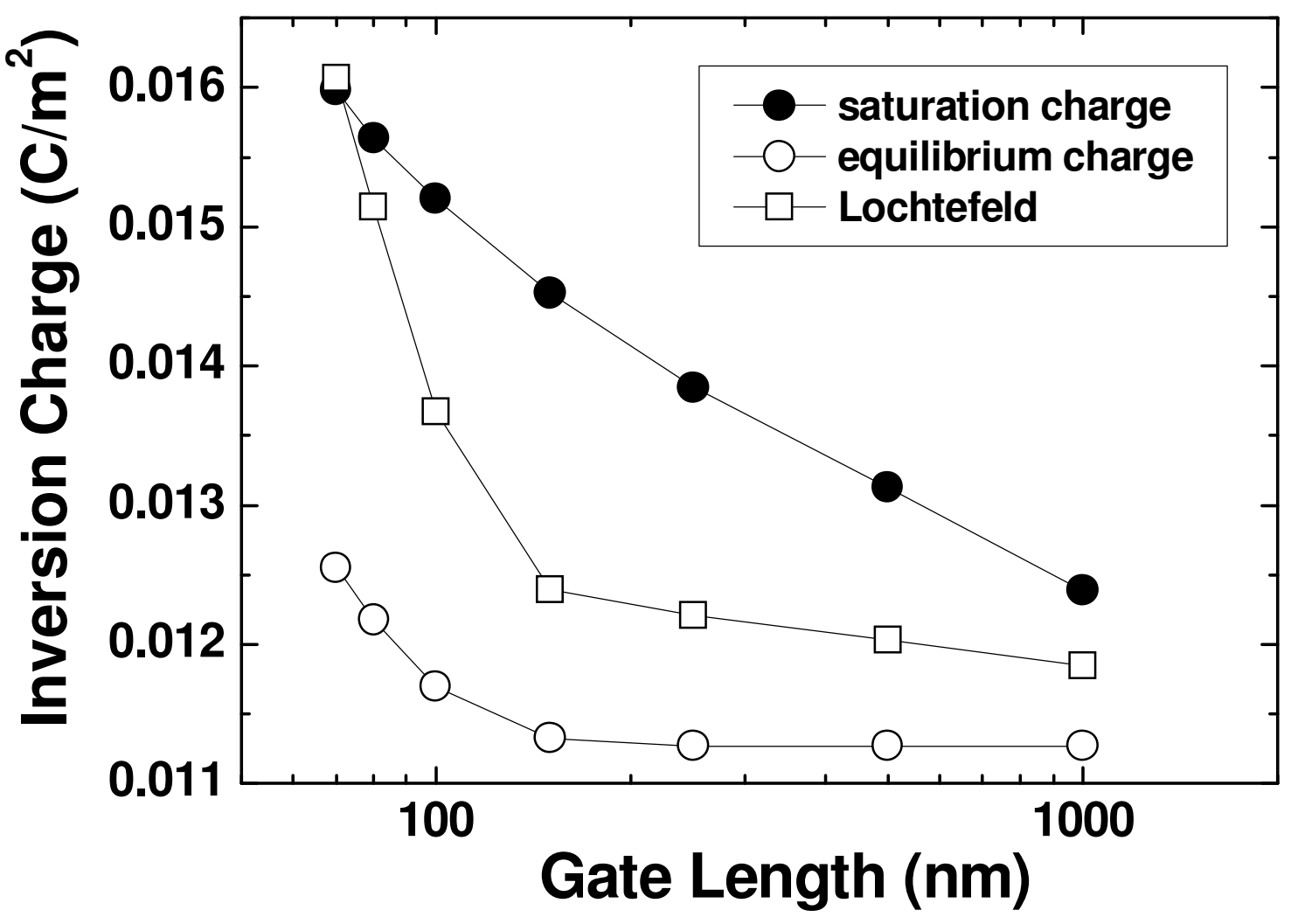

Figure 5